# Computational-design Enabled Wearable and Tunable Metamaterials via Freeform Auxetics for Magnetic Resonance Imaging

*Ke Wu, Xia Zhu, Thomas G. Bifano, Stephan W. Anderson [*], and Xin Zhang [*]*

Dr. K. Wu, X. Zhu, Prof. X. Zhang
Department of Mechanical Engineering, Boston University, Boston, MA 02215, United States.
E-mail: xinz@bu.edu

Dr. S. W. Anderson
Boston University Chobanian & Avedisian School of Medicine, Boston, MA, 02118, United States.
E-mail: sande@bu.edu

Dr. K. Wu, X. Zhu, Prof. T. G. Bifano, Dr. S. W. Anderson, Prof. X. Zhang
Photonics Center, Boston University, Boston, MA 02215, USA.
E-mail: xinz@bu.edu

**Abstract**

Metamaterials hold significant promise for enhancing the imaging capabilities of MRI machines as an additive technology, due to their unique ability to enhance local magnetic fields. However, despite their potential, the metamaterials reported in the context of MRI applications have often been impractical. This impracticality arises from their predominantly flat configurations and their susceptibility to shifts in resonance frequencies, preventing them from realizing their optimal performance. Here, we introduce a computational method for designing wearable and tunable metamaterials via freeform auxetics. The proposed computational-design tools yield an approach to solving the complex circle packing problems in an interactive and efficient manner, thus facilitating the development of deployable metamaterials configured in freeform shapes. With such tools, the developed metamaterials may readily conform to a patient's kneecap, ankle, head, or any part of the body in need of imaging, and while ensuring an optimal resonance frequency, thereby paving the way for the widespread adoption of metamaterials in clinical MRI applications.

**Introduction:**

Metamaterials, constructed by assemblies of multiple judiciously designed structures at subwavelength scale, have emerged as a powerful tool to tailor the effective properties of materials by manipulating the amplitude, phase, and polarization of propagating waves, realizing exotic phenomena that extend beyond those of conventional materials.[1-3] Electromagnetic (EM) metamaterial-enabled technologies are now shifting towards facilitating a wide range of practical applications from the microwave to the optical regime, such as perfect lenses,[4] invisible cloaks,[5] absorbers,[6] holography,[7] and antennas,[8] to name a few. One notable property of metamaterials is near-field enhancement. When excited by an incident wave at their resonance frequency, metamaterials generate an intense and localized electric field at the edges of their narrow capacitive gaps, as well as a magnetic field around the conductive or metallic structures, which has enabled their application to phase conjugation,[9] second-harmonic generation,[10] high-sensitivity sensing,[11] and wireless power transfer,[12,13] among others. Leveraging this unique magnetic field enhancement property, metamaterials, consisting of diverse arrays of unit cells featuring conducting "Swiss rolls", parallel metallic wires, or helical coils, have been utilized to boost the signal-to-noise ratio (SNR) of the magnetic resonance imaging (MRI) by amplifying the radiofrequency (RF) magnetic field strength.[14-20]. However, most of the reported metamaterials for enhancing MRI systems have been constructed in rigid planar configuration, which limits the benefits of near-field enhancement when imaging curved surfaces, such as the brain, neck, or musculoskeletal system (knee, ankle, etc.), as the SNR gains of metamaterials decays rapidly as a function of distance from the metamaterial surface. In addition, metamaterials are susceptible to their local environments and the presence of materials with different permittivities, thus, a precise match in working frequency between the metamaterial and MRI system is challenging when scanning different patients of varying body composition (differing degrees of water, fat, muscle, or bone) in MRI.

To achieve the optimal performance of metamaterials in MRI, previously, we proposed a tunable 'helmet' metamaterial inspired by auxetics for brain imaging.[21] The 'helmet' metamaterial is composed of an array of unit cells featuring metallic helices, the coupling of which leads to a synergy and a collectively resonating mode. When the frequency of the resonant mode approximates the resonance frequency of the MRI system, marked gains in incident RF magnetic fields are achieved, ultimately leading to gains in SNR. The application of auxetics in the 'helmet' metamaterial design refers to a novel class of materials and structures exhibiting a counter-intuitive deformation characterized by a negative Poisson's ratio.[22]. Their unique mechanical properties offer auxetics broad potential applications in diverse engineering, sports, and medical problems.[23-27] This 'helmet', constructed in semi-spherical configuration, ensured a conformal approximation between the metamaterial and human head. More importantly, the unit cells of the metamaterial are tessellated through angulated scissor linkages, forming a deployable auxetic grid, the auxetic isokinetic behavior of which gives rise to the frequency tunability by modulating the coupling between neighboring unit cells. Compared with these commonly adopted tuning mechanisms in tunable metamaterials, which incorporate active materials and are modulated by external influences or signals,[28-30] or tuned by physical perturbation of metamaterials' structures,[31,32] the integration between mechanical auxetic structures and electromagnetic metamaterials offers a novel and straightforward pathway to achieve tunable EM properties without introducing additive materials or extra controlling units. However, most anatomic shapes are geometrically complex and often exhibit strong irregularities and asymmetries. As a result, despite the superiority the method offers, when it comes to the wearable metamaterials configured in a conformal approximation with the surface of the human body parts, the design of deployable freeform auxetics and the corresponding tunable metamaterials presents a significant challenge through manual calculation and construction.

To design a deployable auxetic grid configured in freeform shapes, one feasible strategy is to find its circle packing (CP) pattern for this given geometry, meaning that circles are arranged on its surface such that no overlapping occurs, and no circle can be enlarged without creating an overlap.[33,34]. With the CP pattern, the dimensions and coordinators of the linkages are easily derived. Thus, the design of wearable metamaterials is translated to develop algorithms to find the CP pattern for the given freeform surfaces.[35] However, acquiring the CP patterns for freeform surfaces is a challenging problem in the fields of mathematics and discrete differential geometry. To ease the design process, we propose a computational method aided by a parametric design tool to achieve the CP patterns for any given freeform surface, allowing for the design of deployable auxetic structures, thereby realizing wearable and tunable metamaterials for MRI by integrating the magnetic metamaterials to auxetic structures.

## 2. Results
### 2.1 Circle Packing

To arrange tangent circles on surfaces, a novel type of triangular circle packing mesh (TCPM) was introduced, where the incircles of the triangles form a cohesive packing.[36] In the TCPM, consider two adjacent triangles with vertices V1, V2, V3, and V4. The edges, such as L23 connecting vertices V2 and V3, shared by these two triangles, are tangent to their respective incircles at the contact point P23, as illustrated in **Figure 1**a. To fulfill these geometric properties, the edge lengths of the two triangles must satisfy the following relations:

$$L_{12} + L_{34} = L_{13} + L_{24} \qquad (1)$$

However, the incircles (depicted in orange in Figure 1a) could not form a compact circle packing due to the large gaps around their central vertices. To achieve a more even distribution of gaps among the circles, the contact points of the incircles around a vertex could be utilized to define circles. This set of circles generates a more compact circle packing, represented by the blue circles in Figure 1a. Generating such TCPMs often poses a complex mathematical challenge, involving multiple, and sometimes endless, iterations. Leveraging digital parametric design tools such as the 3D modeling software Rhinoceros® and its plug-in Grasshopper, the optimization algorithm can be solved in an intuitive and interactive environment. This is achieved by manipulating the input parameters with given surfaces, facilitating a more efficient and user-friendly approach.[37] The general optimization process for the TCPM begins with generating an initial triangle mesh that approximates the given surface. The density of the mesh can be adjusted using input parameters such as the mesh edge length, which determines the number of circles in the resulting packing pattern. Subsequently, the surface boundary condition, along with the vertices and edges of the initial mesh, are extracted. Utilizing these parameters, Kangaroo, a component of Grasshopper, is employed to optimize the triangle mesh. This optimization process adheres to constraints related to the incircle packing property, proximity to the given surface, and adherence to surface boundary conditions. Following the optimization, the desired CP pattern of the input surface is achieved. Subsequently, the coordinates of the packing circles' center points, circle diameters, and tangent points are extracted for the subsequent design of deployable auxetic structures. Of note, it is not possible to get a precise circle packing with this method for an arbitrary surface, since the normal axes of tangent circles in a precise packing either pass through a common point or are parallel. Even so, it is still practically useful for general engineering practice. Using the TCPM, it becomes feasible to design freeform auxetics for surfaces with gradual curvature. In this work, we adopted planar shapes with freeform boundaries and hemispherical surfaces as examples to validate this computational method. The circle packing results are depicted in Figures 1b and 1c. For a detailed description of the optimization procedure based on the TCPM, refer to Section S1 and Figures S1 and S2 in the Supporting Information.

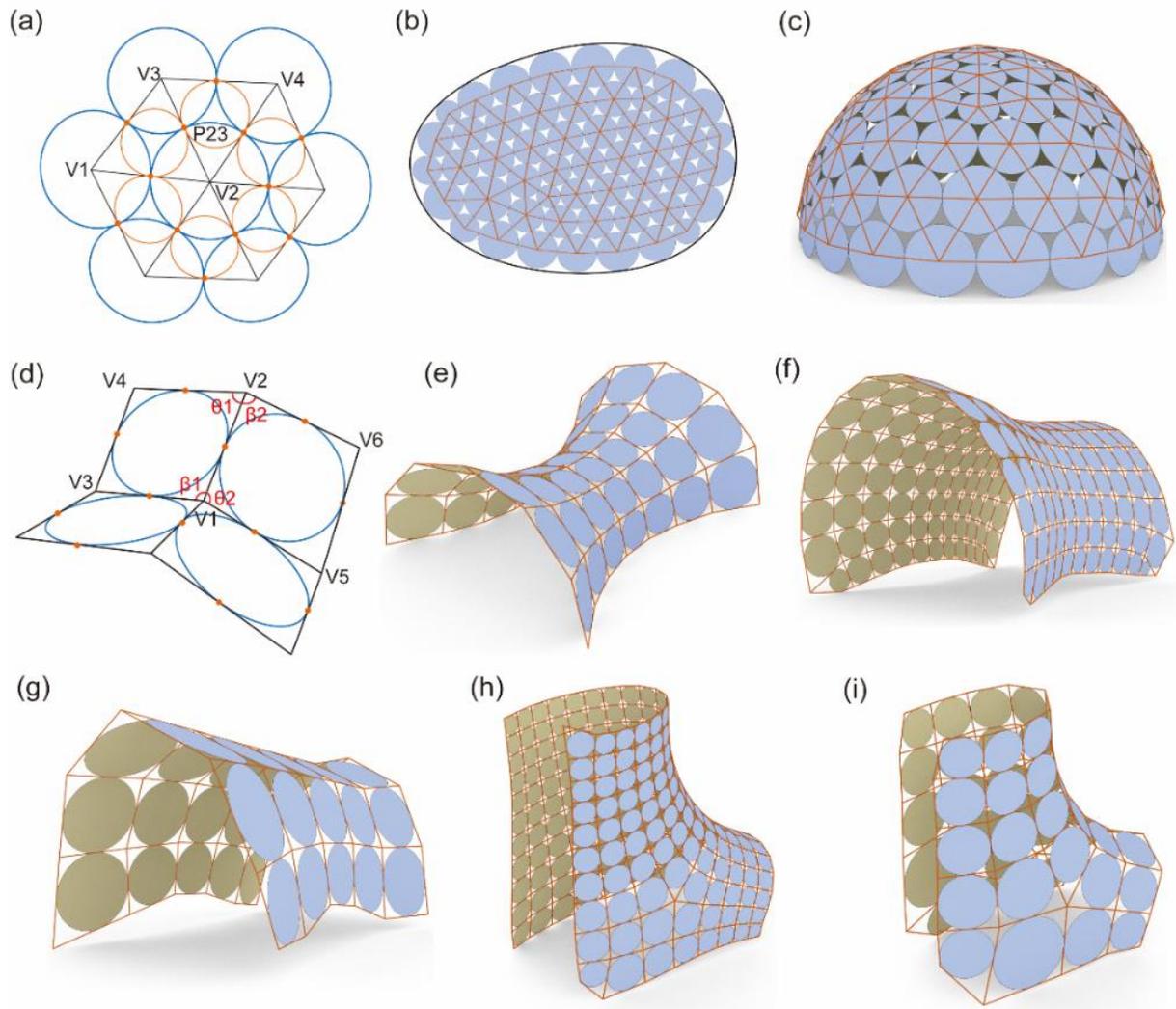

**Figure 1. Circle packing results.** a) Circle pattern formed by the TCPM. b, c) Circle patterns derived from TCPM for planar shape with freeform boundary (b), and hemispherical surface (c). d) Circle pattern formed by the quadrilateral CP mesh. e) Circle pattern derived from quadrilateral CP mesh for the surface of revolution. f–g) Circle patterns for surface configurated in knee shape with different packing densities. h–i) Circle patterns for surface configurated in ankle shape with different packing densities.

When dealing with surfaces exhibiting relative irregularities and large curvatures, the method based on the TCPM may not be practical for achieving a compact and highly tangential circle packing pattern. As an alternative approach, a new type of quadrilateral circle packing mesh (QCPM) for isothermic surfaces was introduced as a strategy to address challenges in discrete differential geometry.[38,39] In these QCPMs, all the quadrilaterals are planar, each face possesses incircles, and the incircles of adjacent quadrilaterals are in contact, as demonstrated in Figure 1d. These geometric properties can be mathematically formulated through the expressions provided below:

$$\begin{cases} \overrightarrow{V_{21}} \cdot \left(\overrightarrow{V_{31}} \times \overrightarrow{V_{41}}\right) = 0 \\ L_{34} + L_{12} = L_{13} + L_{24} \\ \cot\frac{\angle\theta_1}{2}\cot\frac{\angle\theta_2}{2} = \cot\frac{\angle\beta_1}{2}\cot\frac{\angle\beta_2}{2} \end{cases} \quad (2)$$

Likewise, the optimization process for realizing QCPM can be computed in Rhino by adjusting the edge lengths and relocating vertices to meet geometric restrictions, allowing for circle packing on a given surface. To validate this computational method, we utilized a surface of revolution, a subclass of isometric surfaces, as an example. The result is an almost perfect, compact, and highly tangent circle packing, as illustrated in Figure 1e. When dealing with nonisothermic freeform surfaces, the constraints on boundaries and surface shape can be appropriately relaxed to achieve highly tangent packing circles. Considering the configurations of metamaterials for MRI, the focus herein, freeform surfaces modeled from the human knee and ankle, which represent some of the most irregular surfaces on the human body, were utilized as examples to validate the code for computing circle packing. The CP results for the knee are illustrated in Figures 1f and 1g, while the results for the ankle are shown in Figures 1h and 1i (refer to Section S2, Figures S3 and S4 in the Supporting Information). By manipulating the input parameters, packing circles with varying densities could be easily obtained, facilitating the subsequent design and fabrication of wearable metamaterials.

**2.2 Kinematic Study and Fabrication Results**

Once the CP mesh is obtained, the next step is to populate the deployable auxetic structure through angulated scissor linkages and hubs.[40] The angulated scissor linkages consist of a pair of identical kinked rods connected at an intermediate point by a pivot hinge, enabling a relative rotation of the bars about a single axis. Hubs are used as connection joints, linking the endpoints of multiple scissor linkages in a three-dimensional configuration. Additionally, the hubs serve as scaffolding for mounting the metallic helical coil resonators (HCRs) of the metamaterials' unit cells. As an example of dimensioning the angulated scissor units, a portion of the CP results containing 4 circles is extracted and depicted in **Figure 2**a. The dimensions of the kinked rods (as shown at the bottom of Figure 2a) used to connect neighboring hubs can be determined using the coordinates of the circle center points and tangent points, calculated as follows:

$$\begin{cases} l_1 = \lambda_1 |\overrightarrow{T_{12}O_1}| \\ l_2 = \lambda_1 |\overrightarrow{T_{12}O_2}| \\ \Psi_{12} = arccos\left(\dfrac{\overrightarrow{T_{12}O_1} \cdot \overrightarrow{T_{12}O_2}}{|\overrightarrow{T_{12}O_1}||\overrightarrow{T_{12}O_2}|}\right) \end{cases} \quad (3)$$

in which, $\lambda_1$ is an arbitrary number deciding the overall size of the auxetic structure, $l_1$, $l_2$, and $\Psi_{12}$ are the geometric parameters illustrated in Figure 2a. Figure 2b provides an illustrative example of how neighboring hubs and their interconnecting angulated linkages are assembled. This illustration also showcases their kinematic behaviors, enabling stress-free deployment. The $m_i$ in Figure 2b represent the offset lengths reserved for the connection between hubs and linkages. For instance, $m_1$ is given by,

$$m_1 = \lambda_2 |\overrightarrow{T_{12}O_1}| \quad (4)$$

where $\lambda_2$ is an arbitrary number related to the size of hubs. Both $\lambda_1$ and $\lambda_2$ must be consistent for every linkage and hub to ensure geometric compatibilities during deployment. Since the kinked rods embrace a constant angle $\Psi$, all the connected angulated scissor linkages show the same kinematic behavior in a synchronized manner. Consequently, the overall shape of the populated auxetic structure remains fixed, with only its scale changing according to the deployment angle θ. Figure 2c provides a magnified illustration of unit cells in the wearable metamaterials, in which the HCRs are affixed onto the bottom hubs, serving as electromagnetic components of the metamaterial to enhance the magnetic field, thereby amplifying the SNR of MRI. HCRs offer several advantages due to their unique configuration, including their compact sizes, high Q values, and ease of fabrication. One notable property of HCRs is their ability to have their resonant frequency easily tuned by adjusting their geometrical configurations. This flexibility in designing their resonance frequency is particularly crucial in MRI applications where precise frequency control is necessary. To achieve optimal performance in SNR enhancement for wearable metamaterials, the optimization of HCRs plays a pivotal role in the design of the metamaterials (refer to Section S3 and Figure S5 in the Supporting Information). Indeed, the configuration doesn't necessarily have to be coaxial between the HCRs and the hubs. Appropriate axis shifts can result in a more uniform distribution of the resonators during deployment. The metamaterial's resonance frequency tunability is achieved by adjusting the separation distance between resonators, thereby modulating the coupling coefficient through the deployment of the auxetic structure. The relative distance between two hubs as a function of the deployment angle θ can be expressed as:

$$Dis \approx m_1 + m_2 + \sqrt{l_1^2 + l_2^2 - 2l_1 l_2 cos\theta_1} \quad (5)$$

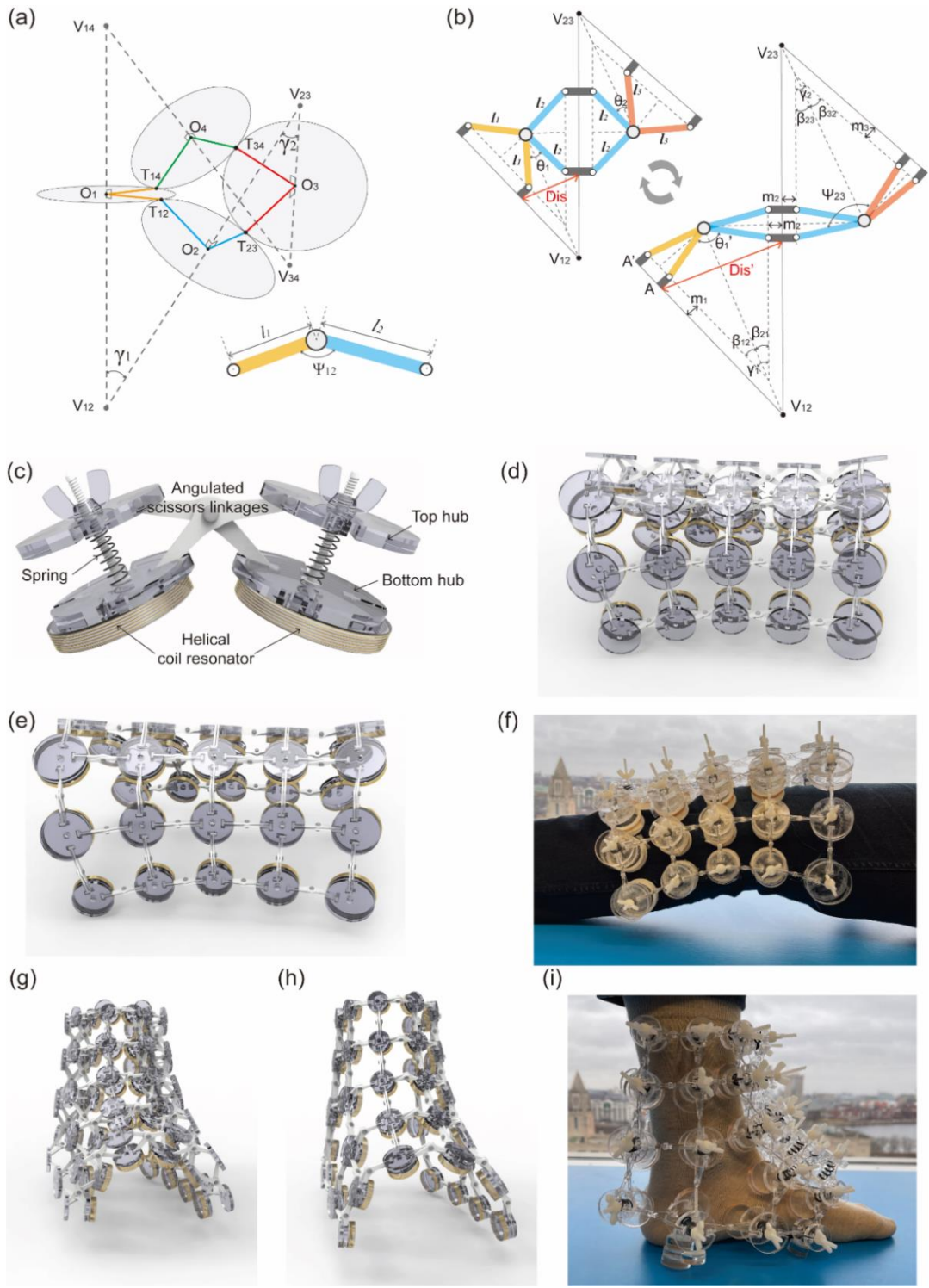

**Figure 2. Design principle for freeform metamaterials.** a) Populating four tangent circles $O_1$ to $O_4$ with angulated scissor units. b) Kinematic behavior of three adjacent tangent circles $O_1$ to $O_3$ shown in (a). c) Magnified conceptual image of inter-connected unit cells of deployable metamaterials. d-e) Illustration of the knee metamaterial with contraction and expansion configurations. g-h) Illustration of the ankle metamaterial with contraction and expansion configurations. Photograph of a human wearing the proposed knee metamaterial (f) and ankle metamaterial (i).

With the design principle and assemble manner reported herein, we successfully fabricated wearable knee and ankle metamaterials for MRI based on the circle packing results depicted in Figures 1g and 1i. Figures 2d and 2e showcase the designed knee metamaterial, demonstrating the reconfiguration of meta-atoms through the auxetic deployment. In Figure 2f, the knee metamaterial is shown being worn by a human subject in a conformal fashion. The designed ankle metamaterial in a contraction and expansion configurations are depicted in Figures 2g and 2h, respectively. A photograph of a human subject wearing the ankle metamaterial, demonstrating a conformal fit between the metamaterial and the surface of the human ankle, is illustrated in Figure 2i. For a detailed visual representation of the deploying process between contraction and expansion states for auxetics configured to approximate knee and ankle shapes, refer to Movies S1 and S2 in the Supporting Information.

**2.3 Electromagnetic Characterization**

Since the unit cells in the knee and ankle metamaterials are irregularly and asymmetrically distributed, we initiated our study by examining a pair of helical coil resonators to investigate their coupling coefficients, resonance modes, and magnetic field distributions. The analytical insights gained from this investigation can be extended to the freeform metamaterials for their EM characterizations. The frequency tunability mechanism of the metamaterial is rooted in the manipulation of the coupling coefficient between the HCRs. This coupling coefficient $k$ can be expressed as follows:

$$k = k_L + k_C = \frac{L_m}{L_s} + \frac{C_m}{C_s} \quad (6)$$

in which $C_m$ and $C_s$ is the mutual capacitance and self-capacitance, and $L_m$ and $L_s$ represent the mutual inductance and self-inductance, respectively. The total coupling coefficient $k$ between two resonators, with an inclined angle of 30º between their axes (as shown in Figure 2c), as well as the contributions from capacitance coupling $k_C$ and inductance coupling $k_L$ are plotted in **Figure 3**a as a function of separation distance. With the interunit cell coupling coefficient, the resonant modes of these two resonators may be derived by employing the coupled mode theory and solving the following equation system:[41]

$$j\omega \begin{bmatrix} a_1 \\ a_2 \end{bmatrix} = j \begin{bmatrix} \omega_1 + j\left(\frac{1}{\tau_{e1}} + \frac{1}{\tau_{o1}}\right) & k\omega_1/2 \\ k\omega_2/2 & \omega_2 + j\left(\frac{1}{\tau_{e2}} + \frac{1}{\tau_{o2}}\right) \end{bmatrix} \begin{bmatrix} a_1 \\ a_2 \end{bmatrix} + \begin{bmatrix} \sqrt{\frac{2}{\tau_{e1}}} \\ \sqrt{\frac{2}{\tau_{e2}}} \end{bmatrix} s_+ \quad (7)$$

in which the subscripts '1' and '2' indicate the two resonators. $a_n$ ( where n=1, 2) represents the mode amplitude of the resonator, $(1/\tau_{en}+1/\tau_{0n})$ denotes the decay rates of the oscillating strength of the resonators due to radiation and intrinsic losses, $\omega_n$ represents the resonance frequency, $s_+$ is a harmonic excitation signal function with frequency $\omega$ (i.e., $s_+=|s_+|e^{j\omega t}$), and $\sqrt{2/\tau_{en}}$ is the coefficient expressing the degree of coupling between the resonator and the excitation signal. Finally, given the mode amplitudes of these two resonators, the reflection spectrum of the array can be expressed by:[41]

$$r = -1 + \frac{\sqrt{\frac{2}{\tau_{e1}}}a_1 + \sqrt{\frac{2}{\tau_{e2}}}a_2}{2|s_+|} \qquad (8)$$

The theoretical reflection spectra with varying coupling coefficients are illustrated in Figure 3b. The detailed mathematical modeling process for charactering a pair of resonators is described in Section S4, Figure S6, Table S1 of the Supporting Information. Within this analysis, two distinct resonant modes emerge as discrete dips on the plotted curves for the two-unit array. In the resonance modes at lower frequency, the induced electric currents circulating along the helical coils exhibit opposite directions. This configuration results in the cancellation of their corresponding magnetic fields, as depicted in Figure 3c. In contrast, the higher frequency mode showcases identical electric currents, leading to the superposition of their induced magnetic fields. This phenomenon significantly enhances the excitation signal, as illustrated in Figure 3d. Consequently, the resonant mode where the induced field are enhanced should be referred to as the working mode for the MRI. Moreover, upon comparing the spectra with varying coupling coefficients, as depicted in Figure 3b, it becomes evident that adjusting the coupling coefficient offers a means to tune the frequency of the working resonance mode within the coil resonator array.

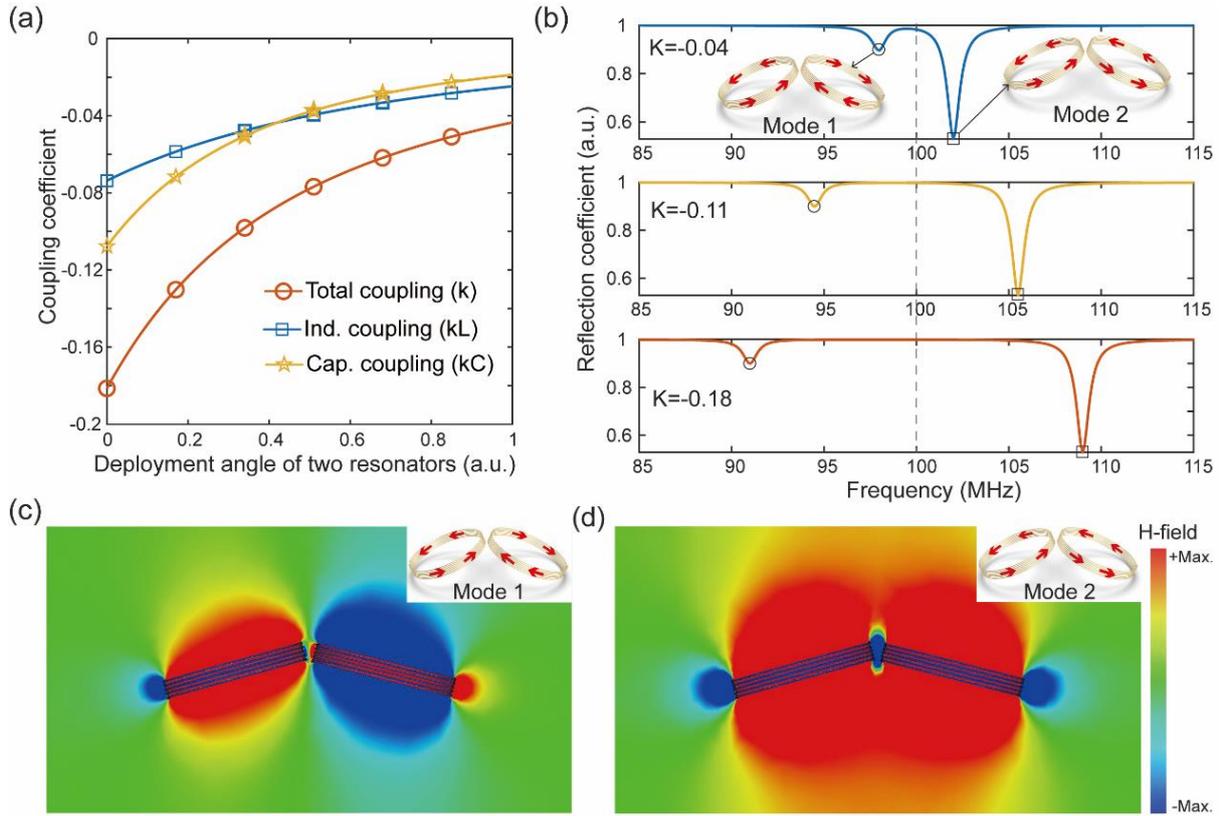

**Figure 3. EM characterization for a pair of resonators.** a) Coupling coefficients as a function of deployment angle. b) Reflection spectrum with different coupling coefficients, e.g., k = -0.04, k= -0.11; and k = -0,18; respectively. c-d) Magnetic field distribution at the resonance mode where the induced current along opposite and identical directions, as shown in the inset figures.

With the mechanism of frequency tunability and magnetic field distribution described above, we characterized the EM properties of the knee and ankle metamaterials for MRI applications, as shown in Figure 2f and 2i. Initially, the reflection spectra of the knee metamaterial were meticulously analyzed using numerical simulations conducted via CST Microwave Studio. Subsequently, the results were validated through experimental testing utilizing a network analyzer, with the outcomes graphically represented in **Figure 4**a. Multiple resonance modes were observed in the knee metamaterial, identifiable as dips on the plotted reflection spectrum. Among these, the working mode resonates at the highest frequency, where the direction of the electric current is uniform across each coil. The magnetic field distribution at this working mode is depicted in Figure 4b, highlighting a substantial enhancement of the magnetic field. Additionally, we extracted the working mode frequencies of the knee metamaterial as it transitioned from contracted to expanded states. Both simulation and experimental results are illustrated in Figure 4c, demonstrating that a ~6 MHz frequency tunability of the working mode may be achieved across the mechanical reconfiguration. This frequency tuning range is sufficient to compensate for detuning effects during imaging and, thereby, ensure an optimized frequency match between the metamaterial and the MRI system.[21]

In addition to the knee metamaterial, the EM properties of the ankle metamaterial were also thoroughly characterized. It is noteworthy that due to the highly irregular and asymmetric distribution of unit cells in the ankle metamaterial caused by the significant curvature of the ankle shape, exciting the working mode, where induced currents in each resonator create a superimposed and enhanced magnetic field inside the auxetic structure, proves to be challenging. To mitigate this issue, a solution was implemented by attaching the helical coil resonators only to the hubs on the left and right sides of the auxetic structure, as opposed to every bottom hub. Figure 4d illustrates both the experimental and simulated reflection spectrum of the ankle metamaterial, highlighting two distinct resonance modes indicated by dips on the curve. To analyze the magnetic field patterns at these two modes, simulations were conducted to visualize the magnetic field distribution on the metamaterial cross-section, represented by the blue plane in the inset of Figure 4f. As depicted in Figure 4e, the left image shows the magnetic field pattern at the working mode, where the induced electric current generates an enhanced magnetic field. In contrast, the right image displays the field pattern at resonance mode 2, where the magnetic field inside the metamaterial weakens due to the cancellation effect between the resonators on the left and right sides of the metamaterial. Consequently, it can be concluded that the left dip on the reflection spectrum in Figure 4d represents the working mode of the ankle metamaterial for MRI applications. Furthermore, like the knee metamaterial, the resonance frequency tunability of the ankle metamaterial was investigated. The experimental and simulated results, shown in Figure 4f, demonstrate a tuning range of approximately 6 MHz achieved through the deployment of the ankle metamaterial. Detailed setups for the experimental tests of the reflection spectra for both knee and ankle metamaterials can be found in Figure S7 of the Supplementary Information.

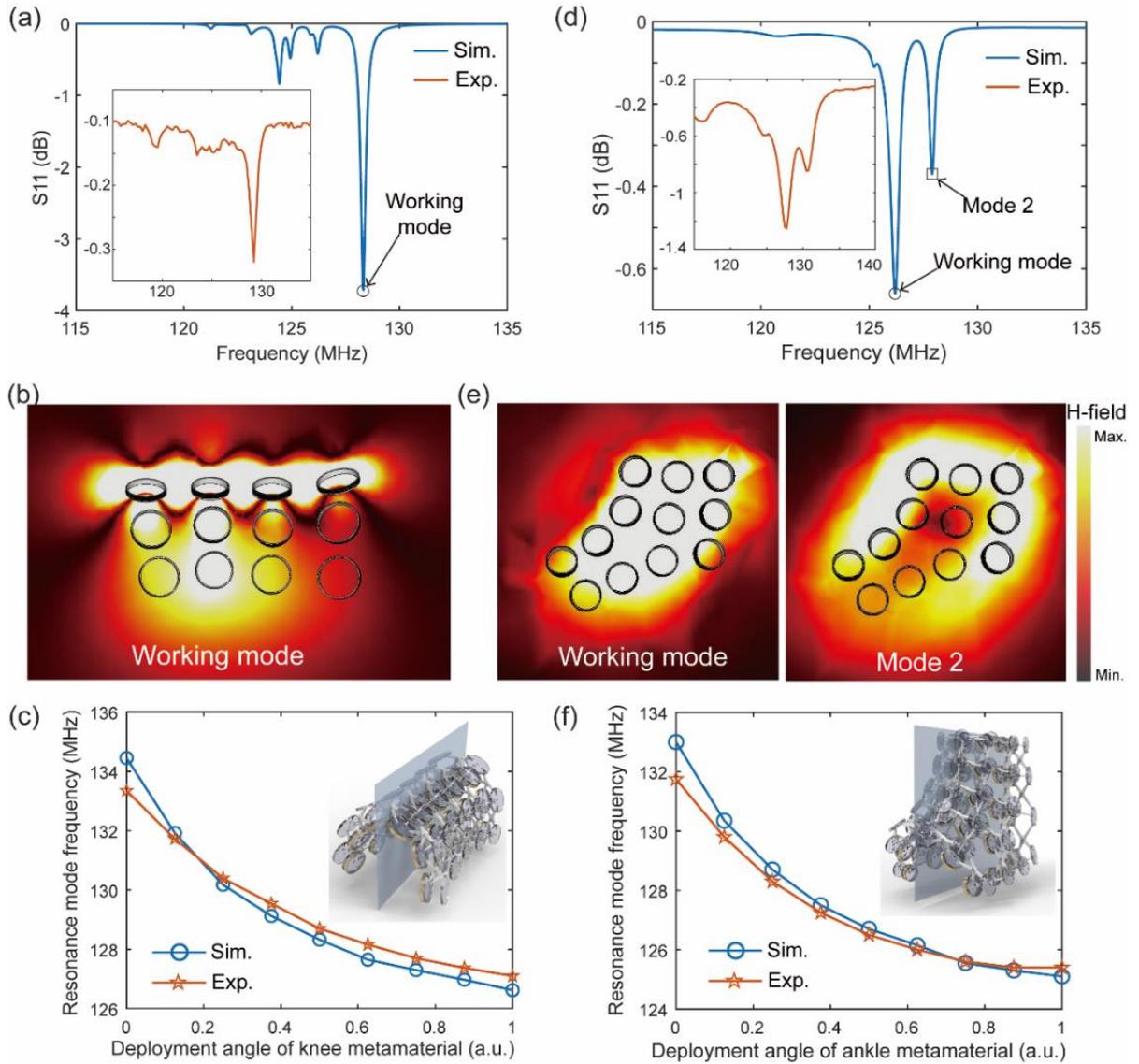

**Figure 4. EM characterization for knee and ankle metamaterials.** a,d) Reflection spectrum of the knee (a) and ankle (d) metamaterials. b) Magnetic field distribution on the metamaterial cross-section (depicted as the blue plane in the inset of (c)) at working mode. c,f) Resonance frequency tunability of the knee (c) and ankle (f) metamaterials as a function of deployment angle. e) Magnetic field distribution on the metamaterial cross-section (depicted as the blue plane in the inset of (f)) at different resonance modes.

## 3. MRI validation

With the EM characterization, we performed experiments in a 3T clinical MRI for these wearable knee and ankle metamaterials to validate their performance in boosting SNR of the images. To demonstrate their conformability, we fabricated two home-made phantoms which are configured in the knee and ankle shapes, respectively. The two-image method was employed to evaluate the SNR values for the MRI images,[42] in which an image of the phantom is acquired using the gradient echo imaging, and an image of noise is captured by turning off the transmission RF coil (see these two images in Figure S8, Support Information). The SNR of the phantom image was referred as the ratio between the mean value of the phantom image magnitude and the standard deviation of the noise image. The knee-shaped phantom filled with 1% agarose gel was scanned with the gradient echo imaging sequence by the body coil (BC) in the absence of metamaterial, which serves as a reference standard for the following SNR comparisons. Next, with the phantom in its original position, and placing the knee metamaterial over the top surface of the phantom with the separation distance between the metamaterial and the top surface of phantom of approximately 20 mm, the experimental setup is shown in **Figure 5**a. By exploiting the tunability of the metamaterial, the resonance frequency of the metamaterial can be precisely tuned to its working mode by adjusting the deployment angle of the auxetic structure. This capability ensures a precise frequency match between the metamaterial and the MRI system. The resulting SNR images of the phantom, obtained both in the absence and presence of the metamaterial, are illustrated in Figure 5b and 5c, respectively. The top and bottom images in Figure 5b and 5c represent the sagittal and axial planes of the phantom, respectively. Comparing with reference image's uniform pattern, the metamaterial-enhanced images exhibit increased signal intensity, akin to the color map of the magnetic field pattern depicted in Figure 4b. This correspondence highlights the direct relationship between magnetic field amplification and the subsequent SNR enhancement facilitated by the metamaterial. To facilitate quantitative comparisons, the SNR enhancement ratios were extracted theoretically and experimentally along the dashed lines in the metamaterial-enhanced image (depicted in Figure 5c). These ratios were normalized to the mean SNR value in the reference image and are illustrated in Figure 5d. The results clearly demonstrate a significant ~3-fold increase in SNR when the knee metamaterial is applied to knee imaging in MRI. (The estimation of the theoretical SNR can be found in Section S5 and Figure S9, Support Information.)

To validate the performance of the ankle metamaterial, a similar experimental approach was adopted. An ankle-shaped agarose gel phantom was scanned, and the experimental setup is depicted in Figure 5a. Due to the irregularity of the ankle phantom, SNR images were acquired in sagittal, oblique, coronal, and axial planes. This comprehensive investigation allowed for a thorough comparison of the ankle metamaterial's performance in MRI. The locations of these planes are indicated by the blue sheets in their corresponding inset figures, as shown in Figures 5e-5h. The left images in Figures 5e-5h present the SNR images captured in the absence of the metamaterial, serving as a reference standard. On the right side in Figure 5e-5h, the metamaterial-enhanced SNR images of the ankle phantom are depicted. Unlike the uniform low SNR values throughout the reference images, the SNR in the central areas of these four ankle images, enhanced by the metamaterial, has demonstrably increased. The enhanced pattern of the sagittal image aligns well with the magnetic field colormap. Of importance, the ankle metamaterial exhibits a deep enough penetration depth for field amplification and SNR enhancement, reaching the central area of the ankle phantom. This characteristic holds significant practical implications in clinical MRI, particularly when imaging deeper anatomical structures. Similarly, the normalized SNR enhancement ratio along the dashed line shown in Figure 5e is extracted and plotted in Figure 5i, demonstrating a 4.8-fold increase in SNR (see Section S5 and Figure S9, Support Information).

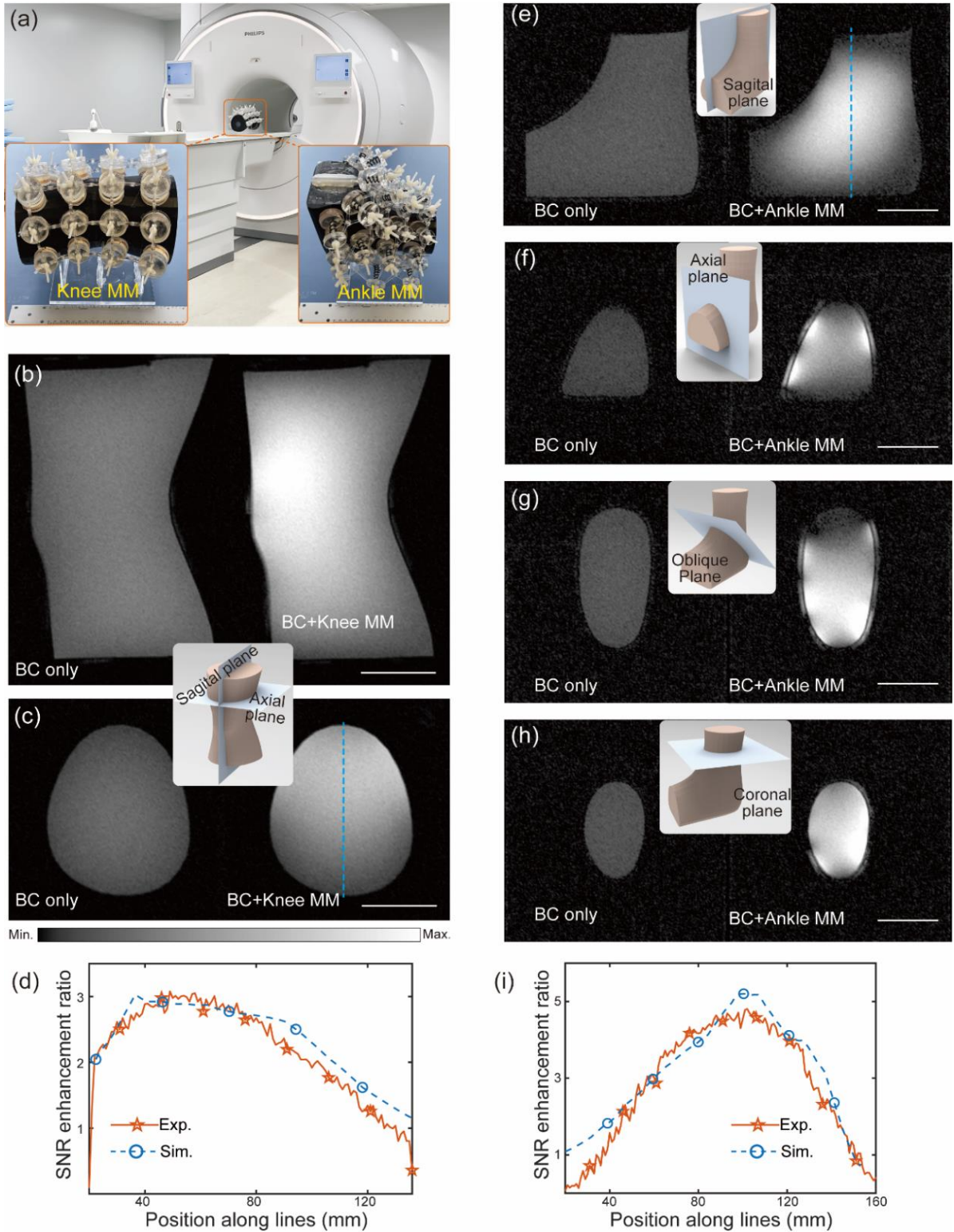

**Figure 5. MRI validations by imaging agarose gel phantoms.** a) Experimental setups. b,c) SNR images on sagittal (b) and axial (c) planes captured by the BC in the absence of knee metamaterial (left image) and in the presence of metamaterial (right image). d, SNR enhancement ratio along blue dashed lines in (c). e-g) SNR images on different cutting planes (indicated by the corresponding inset figures) captured by the BC in the absence/presence of ankle metamaterial. i) SNR enhancement ratio along blue dashed lines in (e).

**Conclusion**

This work demonstrates a computational method for designing the 3D wearable and tunable metamaterials via freeform auxetics for MRI applications. The digital tools reported herein are created in the 3D modelling software Rhinoceros®, offering an interactive and efficient environment to realize the circle packing patterns for freeform surfaces. A design principle to create deployable auxetic structures based on circle packing was combined with a kinematic assembly process to produce wearable metamaterials comprised of helical coil resonators for knee and ankle imaging in MRI. Mathematical modeling based on CMT and experiments were used to characterize the metamaterials' resonance modes, field enhancement, and frequency tunability. Lastly, MRI experiments were performed for knee and ankle metamaterials to validate their performance in boosting SNR of MRI. The deployable metamaterials configured in freeform shapes presented herein improve conformality of the metamaterial to the object of interest and feature tunable resonance frequency thereby taking greater advantage of near-field enhancement of the MRI signal. The approach demonstrated in this work can be extended to other electromagnetic and mechanical metamaterial-based sensing. Crucially, the computational method, driven by digital and parametric design tools, eliminates barriers in solving complex geometrical and structural challenges.

**Methods**

**Circle packing tools.** The parametric design tools were developed in Grasshopper®, which is an auxiliary plug-in component in the 3D modelling software Rhinoceros®. The Grasshopper is a visual block coding language that creates programs by manipulating program elements graphically instead of specifying them textually. The implementation process of circle packing is illustrated in detail in Sections S1, S2, Figures S1-S4 of Support Information.

**Geometry and fabrication of metamaterial.** The reported knee and ankle metamaterials were fabricated by integrating the EM resonators to deployable auxetic structures. The deployable structures are constructed by assembling the well configured angulated scissors linkages and hubs, which are fabricated through laser cut acrylic sheet. The resonators are made from helical copper coil wound around the 3D printed scaffolds with grooves.

**EM Characterization of the metamaterials.** We employed a vector network analyzer (VNA, E5071C, Keysight Inc) with an inductive loop to excite the magnetic resonance of the metamaterials. The reflection spectra S11 were measured and the dips on the curves are correspond to the resonance mode of the metamaterials. The bench test setup for knee and ankle metamaterials are depicted in Figure S7 of Support Information.

**Numerical simulation.** The numerical simulations were performed with CST Microwave Studio software. In the simulation model (Section S5 and Figure S9, Support Information), the dimensions of the metamaterial were the same as the fabricated sample described above.

**MRI validations with phantom.** The BC was employed for both RF transmission and reception. A gradient echo imaging sequence (GRE) was employed using a repetition time (TR) and echo time (TE) of 100 ms and 4.6 ms, respectively. The pixel size was 1 × 1 mm, the slice thickness was 5 mm. GRE imaging was first performed to capture a phantom image (see Figure S8a, Support Information), followed by capturing a noise image by shutting down the transmission RF coil (see Figure S8b, Support Information). The SNR images of the phantom were calculated by the ratio between the mean value of magnitude phantom image and the standard deviation of the noise image.

**Supporting Information**

Support Information is available for this paper.

**Acknowledgements**

This research was supported by the National Institute of Health (NIH) of Biomedical Imaging and Bioengineering Grant No. 1R21EB024673 and the Rajen Kilachand Fund for Integrated Life Science and Engineering. The authors also thank the Boston University Photonics Center for technical support. The authors are grateful to Dr. Yansong Zhao for his experimental assistance during the MRI testing.

**Conflict of Interest**

The authors have filed a patent application on the work described herein, application No.: 16/002,458 and 16/443,126. Applicant: Trustees of Boston University. Inventors: Xin Zhang, Stephan Anderson, Guangwu Duan, and Xiaoguang Zhao. Status: Active.

**Data availability statement**

The data that support the findings of this study are available from the corresponding author upon reasonable request.